\documentclass[%
 reprint,
 amsmath,amssymb,
 aps,
]{revtex4-2}

\usepackage{graphicx}
\usepackage{dcolumn}
\usepackage{bm}

\begin{document}

\preprint{APS/123-QED}

\title{Dark-field and directional dark-field on low coherence X-ray sources with random mask modulations: validation with SAXS anisotropy measurements}

\author{Clara Magnin$ ^{1,2}$}

\author{Laurène Quénot$ ^1$}

\author{Sylvain Bohic$ ^1$}

\author{Dan Mihai Cenda$ ^2$}

\author{Manuel Fernández Martínez$ ^2$}

\author{Blandine Lantz$ ^2$}

\author{Bertrand Faure$ ^2$}

\author{Emmanuel Brun$ ^{1,}$}
\email{Corresponding author: emmanuel.brun@inserm.fr}

\affiliation{$ ^1$Univ. Grenoble Alpes, Strobe Inserm UA7, Grenoble, France
}

\affiliation{$ ^2$Xenocs, Grenoble, France
}%

\begin{abstract}
Phase Contrast Imaging, Dark-Field and Directional Dark-Field imaging are recent X-ray imaging modalities that have been demonstrated to reveal different information and contrast from those provided by regular X-ray imaging. Access to these two types of images is currently limited because the acquisitions require the use of coherent sources such as synchrotron radiation or complicated optical setups. This work demonstrates the possibility of efficiently performing phase contrast, dark-field and directional dark-field imaging on a low-coherence laboratory system equipped with a conventional X-ray tube, using a simple, fast and robust single-mask technique.
\end{abstract}

\maketitle

X-ray Phase Contrast (PC) and Dark-Field (DF) Imaging have been demonstrated to outperform conventional X-ray imaging modalities providing different and complementary information compared to the attenuation-based X-ray imaging \cite{breast, lungDF}. The transfer of phase and dark-field imaging from synchrotron to laboratory equipment is a challenging task \cite{quenot2022} usually employing complex optical setups or highly coherent X-ray sources. 

Concerning DF imaging, initial reports have focused on its development using synchrotron source \cite{pagot2003method,fernandez2002small}. In 2008, Pfeiffer et al. \cite{pfeiffer2008hard} demonstrated the possibility of measuring scattering information about the sample using gratings interferometry (GI) on a conventional X-ray source. This seminal work generated further interest for this technique on a wide range of applications \cite{wang2014non,lungDF, willer2021x, urban2022qualitative,partridge2022enhanced} with innovative optical setups for practical implementations \cite{kagias2021definition,endrizzi2015edge,viermetz2022dark}. A first human clinical trial on lung imaging using the GI method was even made possible at low radiation dose  \cite{willer2021x,urban2022qualitative} showing relevant clinical information about lung injuries not visible using conventional radiography. The adaptation of GI for 3D clinical imaging still has limitations in terms of acquisition time \cite{viermetz2022dark} because of the absorption of the grids used in the optical setup. In addition the method only retrieves the refraction and scattering in one direction, making it insensitive to variations that are parallel to the gratings and causes bad performance in terms of noise for Computed Tomography \cite{kohler2011noise}. At high resolution this latter problem can be solved using circular interferometers \cite{kagias2021definition} but this technique remains restricted to synchrotron so far, to the best of our knowledge. Directional Dark-Field (DDF), sensitive to the direction of the scattering caused by local inhomegenities in the samples, was made possible since the 2010s \cite{jensen2010directional, croughan2022directional} using GI with synchrotron radiation, and in only a few reports using  conventional low-coherence X-ray sources \cite{schaff2017non}. Here, we present the successful implementation of an imaging method allowing one to retrieve the absorption, the phase, the dark-field (DF) and the directional dark-field (DDF) of a sample using a conventional X-ray tube with a simple, fast and robust technique. Moreover we propose in this paper a validation of the DDF information retrieved with our method by Small Angle X-ray Scattering (SAXS) measurements made on the same system.
 Low coherence Modulation Based Imaging (MoBI) method was introduced in a previous publication \cite{quenot2021implicit} demonstrating its efficiency to retrieve phase images on a low coherence X-ray system. Here we extend the MoBI method to DF and DDF images while implementing the technique on a setup that has a SAXS measurement option. The laboratory system used has an even larger X-ray emission spot  than in \cite{quenot2021implicit}  but succeeds in retrieving  quantitative information both in phase and directional dark-field.

\begin{figure}
\includegraphics[width=0.45\textwidth]{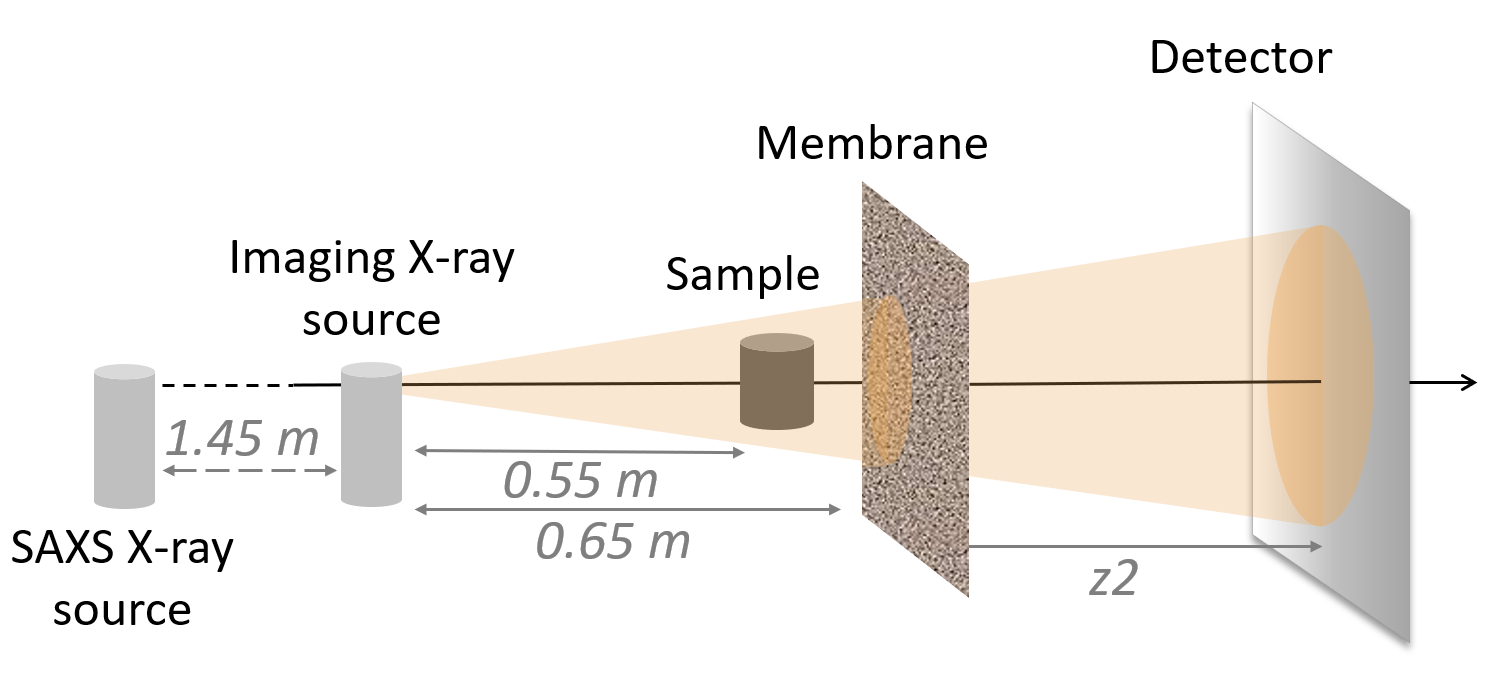}
\caption{Random mask phase-contrast imaging experimental setup: a randomly structured membrane placed in the X-ray beam generate random intensity modulations on the images. A first image $I_r$ is acquired as a reference. Without moving the membrane, a second image $I_s$ is  captured with the sample inserted into the beam. By analyzing the modulation intensity variations in the two images one can retrieve absorption, phase, dark-field and directional dark-field information.}
\label{fig:SetUp}
\end{figure}

The MoBI method is based on the use of granular membranes that allow the X-ray beam to be modulated by the local differences in attenuation in a spatially random manner. In practice, a randomly structured membrane, such as sandpaper is introduced in the path of the beam to produce a first reference image called $I_r$. A second image $I_s$ is then obtained with the sample inserted into the beam (see Fig. \ref{fig:SetUp}) while keeping the membrane static. This operation can be repeated several times while moving the membrane, and then obtaining couples of measurements $I_s$ and $I_r$. The interactions of the photons with the sample generate local distortions of the reference pattern that we want to track. By numerical comparison of the images $I_s$ and $I_r$, one can retrieve absorption, phase, dark-field and directional dark-field information using a digital process, named LCS, described later. The refraction of the sample, linked to the phase, can be measured by looking at small local shifts of the random intensity pattern. The scattering events, which contribute to the dark-field signal, can be tracked by measuring the blurring of the reference pattern.

The experimental setup used for this study is based on a Xeuss 3.0 instrument (Xenocs SAS, Grenoble France) adapted for MoBI. The device is made of two X-ray sources which deliver a monochromatic collimated beam for Small/Wide Angle X-ray Scattering (SAXS/WAXS) measurements and a polychromatic cone-shaped beam dedicated to imaging. A vacuum chamber contains the sample, a random membrane, and a movable detector. The experimental parameters for this study are summarized in Table \ref{tab:ExpPram} and are shown in Fig. \ref{fig:SetUp}.

\begin{table*}
\caption{Experimental setup parameters}
\begin{ruledtabular}
\begin{tabular}{ccccc}
    Distances&SAXS source&Imaging source&Detector
&Membrane\\ \hline

 Source to sample:&Genix3D X-ray tube&X-ray tube  &Photon counting &  \\
 
  0.55 m&Copper anode
 &Copper anode&Eiger2 &\\
 
Sample to membrane: &50 kVp &30 kVp&DECTRIS, Switzerland& Piece of (5 cm * 10 cm)\\

 0.1 m& Monochromatic & 8.6 keV &Pixels of 75 $\mu$m&of sandpaper p180\\
 
 Membrane to detector:& 8 keV (Cu K$\alpha_{1,2}$)& Spot size: &60s per image&(DEXTER PRO)\\

 0.400 m to 4.510 m&Pencil beam 150 $\mu$m   & 50 $\mu$m &600s per SAXS acquisition&\\

\end{tabular}
\label{tab:ExpPram}
\end{ruledtabular}
\end{table*}


The SAXS source is used in this study to validate the DDF retrieval method by comparing the anisotropy of the SAXS signal with the orientation values of the DDF obtained. The SAXS technique consists of measuring the scattering of X-rays by the sample and retrieve information on the structure of the constituent material (size, shape, arrangement, crystallinity) \cite{SAXSdef, jeffries2021small}. It is possible to study elements on a size range from Ångströms to microns. The detected signal is an average over the scattering volume, corresponding to the intersection between the incoming beam (typically 200 $\mu$m in diameter) and the sample. The 2D scattering image is converted to 1D profiles, corresponding to the scattered intensity as a function of the module of the scattering vector or its azimuth.

The first steps of  the theoretical development of the implicit method called Low Coherence System (LCS) has been presented in  \cite{quenot2021implicit}. Here, we extend this approach, by using a Fokker-Planck-like approach \cite{paganin2019x,morgan2019applying}, in order to incorporate dark-field and directional dark-field retrieval. For this purpose we start from the LCS final equation:
\begin{equation}
\label{eq:TIESpeckleLowCoherence}
\begin{aligned}
I_{r}(x,y)-\frac{I_{s}(x,y)}{I_{obj}(x,y)} \approx  D_{\perp}(x,y)\nabla_{\perp}  [I_{r}(x,y)]\\
\approx  D_x(x,y)\frac{\partial I_r(x,y)}{\partial x}+D_y(x,y)\frac{\partial I_r(x,y)}{\partial y}
\end{aligned}
\end{equation}
Where, $D_{\perp}=(D_x, D_y)$ is the transverse displacement field and $\nabla_{\perp}={\partial}/{\partial x}+{\partial}/{\partial y}$ is the two dimensional transverse gradient operator. $I_{obj}$ is a sink term introduce to compensate for attenuation that might comprise also the interference fringes, if any. 
Eq. \ref{eq:TIESpeckleLowCoherence} is then completed with a diffusion term $\nabla_\perp^2[D_f(x,y)I_r(x,y)]$. Assuming small variations of the displacement function the previous term can be simplified as $D_f(x,y)\nabla_\perp^2[I_r(x,y)]$, as done previously in \cite{pavlov2020x}. The gradient of the displacement, which corresponds to the phase Laplacian representing edge-enhancement effects, is weak with a low-coherence source. Consequently, for better results, we experimentally  maximize the gradient of the reference image to extract the pattern displacement caused by the sample  according to this term $D_\perp(x,y)\nabla_\perp[I_r(x,y)]$. 
\\
Finally we get the equation:
\begin{equation}
\begin{aligned}
\label{eq:TIE_LCS}
I_r(x,y)-\frac{I_s(x,y)}{I_{obj}(x,y)}&=D_\perp(x,y)\nabla_\perp[I_r(x,y)]\\&-z_2 D_f(x,y)\nabla_\perp^2[I_r(x,y)]
\end{aligned}
\end{equation}
Where $z_2$ is the sample-to-detector distance.
In order to retrieve the absorption, phase and dark-field signal, we must solve a system with at least 4 membrane positions due to the 4 unknowns ($I_{obj}$, $D_x$, $D_y$, $D_f$):

\begin{equation}
\begin{aligned}
\label{LCSDFsystem}
I_{r}^{(k)}(x,y) &= \frac{1}{I_{obj}(x,y)}I_{s}^{(k)}(x,y)+  D_x(x,y)\frac{\partial I_r^{(k)}(x,y)}{\partial x}\\&+D_y(x,y)\frac{\partial I_r^{(k)}(x,y)}{\partial y}-z_2D_f(x,y)\nabla_\perp^2[I_r^{(k)}(x,y)],
\end{aligned}
\end{equation}

\begin{figure}
\includegraphics[width=0.47\textwidth]{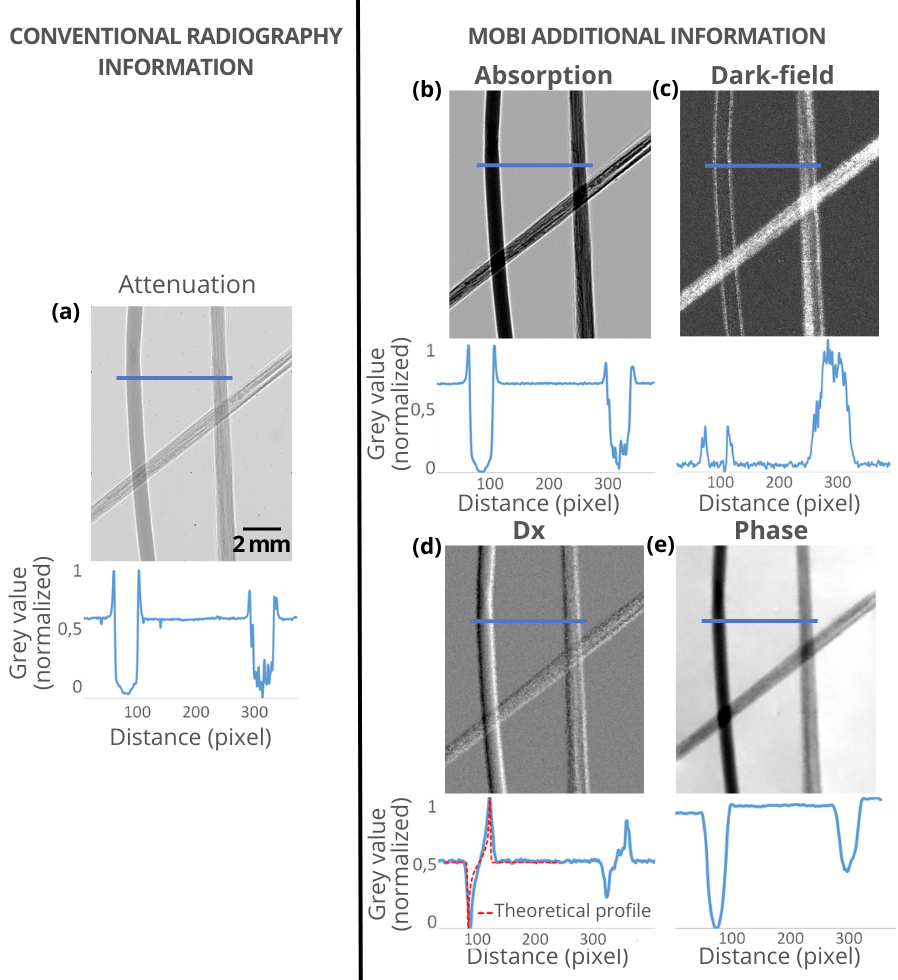}
\caption{Attenuation (a), dark-field (b), dx (c) and phase (d) images of a nylon wire (left) and two bundles of carbon fibers with different orientations. The images are retrieved using the LCS and 10 pairs of reference and sample images at a sample-detector distance of 3200 mm. Their respective profiles along the blue line are given below each image.}
\label{fig:DF}
\end{figure}

Figure \ref{fig:DF} shows the imaging results on a sample composed of a nylon thread of 400 $\mu$m diameter and two bundles of carbon fibers oriented with different angles.
The conventional radiography of the sample in the left sub-Fig. \ref{fig:DF} (a) shows the conventional attenuation of the object made without the use of the membrane. The contrast in this first image is formed by the addition of several physical phenomena: the absorption of photons by the sample but also scattering and phase effects recognizable by the edge effects on the sides of the object despite the low spatial coherence of our system. On the other hand the results of the MoBI method (the right panels in Fig. \ref{fig:DF}) shows the separate contributions of these phenomena into complementary image information: absorption (b), dark-field (c), refraction (d) and phase shift (e). 

In the attenuation (a) and absorption (b) images we can't distinguish nylon and carbon even on the intensity profile due to their similar attenuation properties at this photon energy. However, in the phase contrast image we see differences between the nylon wire  and the  carbon bundles because  nylon generates a stronger phase shift than the fibers. The refraction obtained with our method is in good agreement with the theoretical one calculated with the tabulated values and the thickness of our nylon wire \cite{brennan1992suite}. Similarly, the carbon fibers bundle create a stronger dark-field signal than the nylon thread which highlights the presence of sub-pixel porosity within the bundle, which cannot be resolved by the transmission measurement.

The measured dark-field signal represented in Fig. \ref{fig:DF} (d) can be described as a rank 2 tensor \cite{pavlov2021directional,smith2022x} and then the equation  \ref{LCSDFsystem} becomes:

\begin{align}
  \nonumber I_r(x,y) - I_s(x,y) = & \frac{z_2}{k} \nabla_{\perp} [I_r(x,y) \nabla_{\perp}\phi(x,y)] &&\\ \nonumber &- z_2\frac{\partial^2}{\partial x^2}\left[D^{(xx)}_f(x,y)I_r(x,y)\right]  
  \\ \nonumber &- z_2\frac{\partial^2}{\partial y^2}\left[D^{(yy)}_f(x,y)I_r(x,y)\right]  
  \\  &- z_2\frac{\partial^2}{\partial x\partial y}\left[D^{(xy)}_f(x,y)I_r(x,y)\right]
  \label{eq:TIE_DDF_generalised_form} 
\end{align}

\begin{figure}
\includegraphics[width=0.47\textwidth]{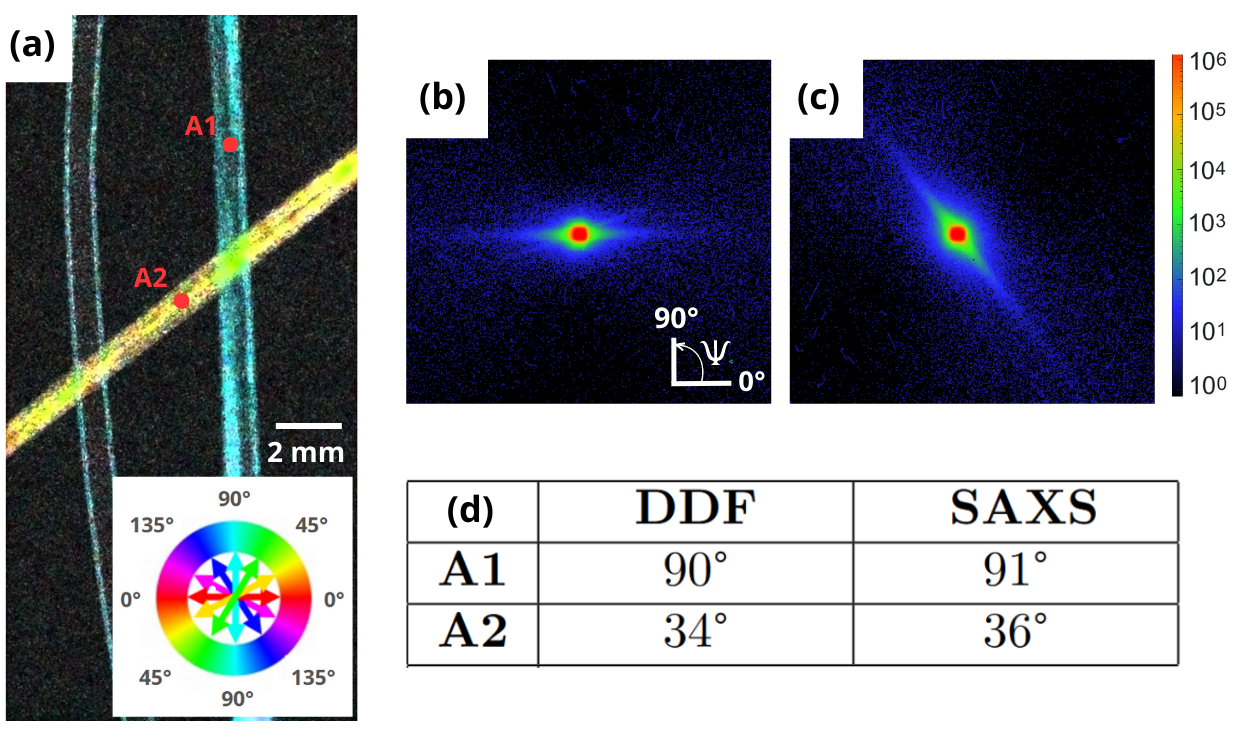}
\caption{Directional Dark-Field image retrieved using the LCS and 10 pairs of reference and sample images at a sample-detector distance of 3200 mm (a). SAXS measurements performed at sample points A1 (b) and A2 (c), marked in red in the image (a). Comparative table of orientation measurements obtained with the two methods (d).}
\label{fig:DDF}
\end{figure}

This tensor can be further decomposed to get the eccentricity of the DF ellipse and its orientation. In many cases, the micro-structures causing the dark-field can be oriented. In our case the bundle of fibers creates a strong scattering signal that is more important in the direction perpendicular to axis of the fibers probably due to multiple refraction. Therefore  the blurring of the pattern measured by our method will have a preferential direction. This hypothesis is confirmed by the experimental measurements shown in Fig. \ref{fig:DDF}. 
The directional dark-field image sub-Fig. \ref{fig:DDF} (a) highlights the two different directions of the two carbon fiber bundles. 
The  carbon fiber orientation retrieved  by DDF was compared to the direction of the scattering pattern obtained by performing SAXS measurements using the monochromatic source of the setup. The details of the angular retrieval technique on SAXS signal is described in supplementary material. The average orientation angle $\Psi$ in SAXS is obtained by solving an anisotropy tensor on an azimuthal angle interval of interest. SAXS was performed at two locations in the sample, shown in sub-Fig. \ref{fig:DDF} (a) by points A1 and A2 which have clearly different orientations in the DDF image. For both methods (DDF and SAXS) the direction probed and shown in sub-Fig \ref{fig:DDF} (d), i.e. the fiber orientation in our case, is perpendicular to the measured scattering signal $\Psi$. The orientation results obtained by SAXS Fig. \ref{fig:DDF} (b) and (c) are really close to the angular values provide by DDF (maximum disagreement of 2 degrees) as shown in sub-Fig. \ref{fig:DDF} (d).

\begin{figure}
\includegraphics[width=0.47\textwidth]{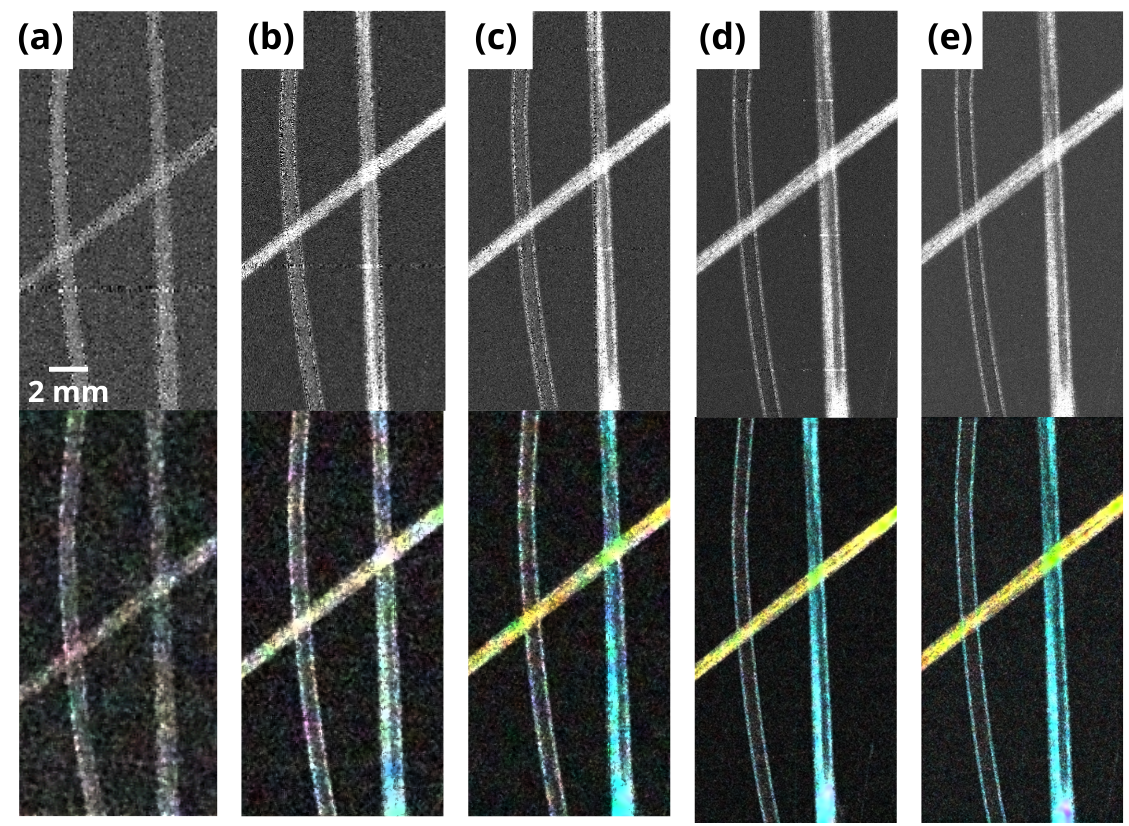}
\caption{Dark-field and directional dark-field images of a nylon wire and carbon fibers retrieved by changing the sample-detector distance. DF (above) and DDF images (below) at a detector distance of 400 mm (a), 800 mm (b), 1200 mm (c), 2200 mm (d) and 3200 mm (e).}
\label{fig:Distance}
\end{figure}

Figure \ref{fig:Distance} shows the dark-field and directional dark-field images according to 5 sample-to-detector distances. To simplify the comparison the images were scaled up to the maximal magnification obtained with the maximal sample-to-detector distance. 

The quality of the dark-field and directional dark-field images is quite stable with the position of the detector for this sample and this setup after a given distance threshold. According to Fig. \ref{fig:Distance}, at a distance of 400 mm the spatial resolution of the system seems to be not sufficient to resolve both DF and DDF even if in this configuration we obtain the maximum photon statistics. At a distance of 800 mm the dark-field information is retrieved but the DDF measurement remains noisy. However, for greater distances, the spatial resolution of the images is improved but the quality of the DF and DDF information remain comparable.
The deterioration of the DF and DDF signals at short distances (sub-Fig. \ref{fig:Distance} (a) and (b)) can be caused not only by the decrease in image resolution but also by the fact that we kept the same random membrane that created, at these distances, smaller modulations not entirely resolved by our numerical system. Indeed, the modulation generated by the membrane in the MoBI method is a key element in the optimization of the quality of the image retrieval. In our device the membrane has been selected to give optimal image quality for a distance of about 2000 mm. Thus, by bringing the detector closer, we change both the magnification of the sample and the pattern of the membrane meaning that the modulation size of the reference is modified and is no longer necessarily optimal.


The development and adaptation of the LCS algorithm, specific to low coherence systems, combined with the use of random absorbing masks, demonstrate the possibility of performing X-ray phase-contrast, dark-field and directional dark-field imaging using a conventional source with a large focal spot. The determination of the average anisotropy direction were quantitative and validated both for phase and dark-field.
The experimental practicality of the setup combined with the simplicity of the numerical processing opens the way to transfer DF and DDF imaging modalities to medical applications. Moreover the method used here is not limited to this particular X-ray energy range and could be applied to higher ranges if necessary.
In future work, the combination of imaging and scattering measurements on the same setup will be used to relate DF and DDF images to SAXS/USAXS and multiple refraction signals.

\section*{Data availability} 
Data underlying the results presented in this paper are available at \href{URL}{https://osf.io/hz2dx}
\\Numerical implementation in python of the phase retrieval is available at \href{URL}{https://github.com/DoctorEmmetBrown/popcorn}

\begin{acknowledgments}
Part of this work was supported by the LABEX PRIMES (ANR-11-LABX-0063) of Université de Lyon, within the program  "Investissements d’Avenir"  (ANR-11-IDEX-0007) operated by the French National Research Agency (ANR). We acknowledge the support of ANRT cifre n°2022/0476 program that partly funds C.M.
\end{acknowledgments}

\bibliography{sample1}

\end{document}